\documentclass[twocolumn,prd,
superscriptaddress,
showpacs,floatfix,preprintnumbers,
nofootinbib, hyperref]{revtex4}

\usepackage{epsfig}
\usepackage{amsmath,bm}

\newcommand{\D}{{\rm d}}
\newcommand{\E}{{\rm e}}
\newcommand{\I}{{\rm i}}

\begin{document}

\title{Adiabatic Faraday effect in a two-level Hamiltonian formalism}

 \author{Basudeb Dasgupta}
 \affiliation{CCAPP, The Ohio State University,
 191 W.~Woodruff Avenue, Columbus, 43210 OH, USA}

 \author{Georg G.~Raffelt}
 \affiliation{Max-Planck-Institut f\"ur Physik
 (Werner-Heisenberg-Institut), F\"ohringer Ring 6, 80805 M\"unchen,
 Germany}

\date{19 June 2010}

\preprint{MPP-2010-62}

\begin{abstract}
The helicity of a photon traversing a magnetized plasma can flip
when the $B$--field along the trajectory slowly reverses. Broderick
and Blandford have recently shown that this intriguing effect can
profoundly change the usual Faraday effect for radio waves. We study
this phenomenon in a formalism analogous to neutrino flavor
oscillations: the evolution is governed by a Schr\"odinger equation
for a two-level system consisting of the two photon helicities. Our
treatment allows for a transparent physical understanding of this
system and its dynamics. In particular, it allows us to investigate
the nature of transitions at intermediate adiabaticities.
\end{abstract}

\pacs{41.20.Jb, 42.25.Lc, 95.85.Sz, 14.60.Pq}

\maketitle

\section{Introduction}                        \label{sec:introduction}

Different polarization states of electromagnetic waves propagating
in media and/or external $B$--fields usually have different
refractive indices, leading to a nontrivial evolution of a given
polarization state. In astronomy, the rotation of the plane of
polarization caused by a magnetized medium along the line of sight
(Faraday effect) is the most important example. Such phenomena are
perfectly analogous to particle oscillations and, in particular, to
neutrino flavor oscillations where the role of polarization is
played by flavor.\footnote{Optical birefringence can be used to
explain neutrino oscillations in a pedagogical demonstration
experiment~\cite{Weinheimer:2010ar}.} Moreover, in addition to
mixing photons of different polarization, $B$--fields also mix
photons with certain other particles, notably gravitons, neutral
pions, or hypothetical axion-like particles, which in this context
play the role of additional photon polarization
states~\cite{Raffelt:1987im}.

If the medium (taken to include external fields) varies along the
photon trajectory, the refractive indices of two polarization states
may cross over. In the adiabatic limit a true crossing is avoided
and opposite helicity states are adiabatically connected, leading to
complete transformation. In neutrino physics, this phenomenon is
known as the Mikheev-Smirnov-Wolfenstein (MSW) effect, allowing for
large flavor transformations even when the vacuum mixing angle is
small~\cite{Wolfenstein:1977ue, Mikheev:1986gs, Bethe:1986ej}. Such
resonant transformation effects have also been studied in the
context of photon--axion oscillations~\cite{Raffelt:1987im}.

However, the case of photons with different polarization has been
ignored until recently when Broderick and
Blandford~\cite{Broderick:2009ix} for the first time considered the
Faraday effect across regions of slow $B$--field reversal where the
dispersion relations of photons with opposite helicities cross over.
If this transition is adiabatic\footnote{Broderick and Blandford use
the term ``super adiabatic'' in this context, but we prefer to keep
the usual terminology.}, photons flip their helicity. The practical
astronomical consequence is that the usual rotation measure for
photon polarization ${\rm RM}\propto\int {\bf B}\cdot\D\bm{\ell}$
changes to $\pm\int |{\bf B}\cdot\D\bm{\ell}|$ and thus can build up
continuously across regions of opposite field direction. The
adiabaticity condition depends on photon frequency. Comparing the
Faraday effect at frequencies above and below a certain $\omega_{\rm
crit}$ may allow one to study the geometry of astrophysical magnetic
fields.

The main purpose of our paper is to formulate the adiabatic Faraday
effect in the more familiar language of particle oscillations using
a linearized wave equation. In this way the evolution of photon
polarization is described by a Schr\"odinger equation for the
two-level system consisting of the two photon helicity states. While
the underlying physics, of course, is precisely as discussed in
Ref.~\cite{Broderick:2009ix}, our approach helps to bring out the
explicit analogy to neutrino oscillations and therefore allows one
to borrow both intuition and results from more familiar cases. In
addition, we recover the nature of Faraday rotation at intermediate
adiabaticity.

The article is organized as follows. In Sec.~\ref{sec:EoM} we derive
a two-level Schr\"odinger equation for photon polarizations and
identify the condition for adiabatic helicity flipping. In
Sec.~\ref{sec:faraday} we work out the impact on linearly polarized
states and recover the adiabatic Faraday effect. We also discuss a
signature for the transition from adiabatic to non-adiabatic
evolution. We conclude in Sec.~\ref{sec:conclusions}.

\section{Hamiltonian Approach}
\label{sec:EoM}

\subsection{Linearizing the wave equation}          \label{sec:linear}

The evolution of photon polarization, neutrino flavor oscillations,
and similar phenomena derive from the underlying wave equation for
the relevant fields. Since dispersion plays a central role, it is
easiest to assume harmonic time variation $\E^{-\I\omega t}$ for all
fields, i.e.\ to study the spatial variation of a monochromatic
wave. Using natural units with $\hbar=c=1$, one finds a stationary
Klein-Gordon equation of the form\footnote{We use sans-serif letters
to denote matrices in polarization space and bold faced sans-serif
letters for ``spinors'' in this space.}
\begin{equation}\label{eq:KG1}
-\bm{\nabla}^2\,\bm{{\sf A}}=
\left({\omega^2}-{\sf\Pi}\right)\bm{{\sf A}}\,,
\end{equation}
where $\bm{{\sf A}}$ is a ``spinor'' of amplitudes describing the
multi-component wave phenomenon. For photon propagation, this is the
Jones vector formed by the transverse components of the electric
field (Appendix~A), whereas for neutrinos the wavefunction in flavor
space comes into play. For photons, ${\sf\Pi}$ is the ``polarization
tensor,'' whereas for neutrinos, ${\sf\Pi}={\sf M}^2$ with ${\sf M}$
being the mass matrix that is non-diagonal in the interaction basis.

The problem simplifies further if the waves are relativistic, i.e.\
$\omega^2\gg |m^2|$, where $m^2$ is a typical eigenvalue of ${\sf
M}^2$ or of ${\sf\Pi}$. In the language of refractive indices, the
dispersion relation is written as $k=n\omega$ and the relativistic
assumption amounts to $|n-1|\ll 1$. The propagating waves now
involve a short length scale (the wave length) and a long one
corresponding to the phase difference (the oscillation length).
Assuming propagation in the $z$--direction, one can easily separate
the fast and slow variation by the nominal
substitution~\cite{Raffelt:1987im}
\begin{equation}
-\left({\partial}_z^2+\omega^2\right)=
(\I\partial_z+\omega)(\I\partial_z-\omega)
\to -2\omega(\I\partial_z+\omega)\;,
\end{equation}
where we have used $-\I\partial_z\to k\approx\omega$ in one of the
terms. This approximation replaces the Klein-Gordon equation with a
linear equation
\begin{equation}
\I\partial_{z}\bm{{\sf A}}=
\left(-\omega+\frac{{\sf\Pi}}{2\omega}\right)\bm{{\sf A}}\,.
\end{equation}
Moreover, we are not interested in the overall phase of the wave,
but only in phase differences between different components.
Therefore, on the r.h.s.\ we may drop terms that are proportional to
the unit matrix and find
\begin{equation}\label{eq:schrodinger}
\I\partial_{z}\bm{{\sf A}}={\sf H}\bm{{\sf A}}\,.
\end{equation}
The ``Hamiltonian'' is ${\sf H}={\sf\Pi}/2\omega={\sf M}^2/2\omega$.

Sometimes Eq.~(\ref{eq:schrodinger}) is written in terms of a
parameter ``time,'' playing the role of an affine parameter along
the trajectory. The linear wave equation then manifestly resembles a
Schr\"odinger equation for a two-level system such as a spin
evolving in a magnetic field. We stress, however, that
Eq.~(\ref{eq:schrodinger}) is a {\em classical\/} equation. Using
natural units somewhat obscures that $\hbar$ does not appear anyway,
at least not when considering the variation of photon polarization
along a trajectory.

The analogy to a Schr\"odinger equation reveals that the linear wave
equation provides for unitary evolution along the beam, i.e., any
photon that disappears from one polarization state must appear in
another. Linearizing the wave equation has removed such physical
effects as reflection by inhomogeneities or discontinuities of the
medium. In the context of neutrino physics such effects are always
vanishingly small, except perhaps near a supernova core, whereas for
photons near propagation threshold, polarization-dependent
reflection effects may play a nontrivial role.

Likewise, the wave may suffer deflection caused by density
variations in the transverse direction. Therefore, the requirement
that all polarization components follow the same trajectory with
sufficient precision poses nontrivial
constraints~\cite{Broderick:2009ix}. Typically, relativistic
propagation will be required because near propagation threshold the
refractive indices between different components are largest and the
differential deflection of different polarization states would be
largest.

The connection between a classical two-level equation and a true
quantum equation as well as the precise role of the relativistic
approximation will be explored elsewhere. Here we simply take
advantage of the formal equivalence of our problem with a two-level
quantum system.

\subsection{Photon dispersion in magnetized plasma}

Photon dispersion in a cold collisionless plasma with electron
density $n_e$ and weak external magnetic field ${\bf B}$ is
determined by the plasma frequency and cyclotron frequency that are
respectively\footnote{In the particle-physics literature,
rationalized units with $\alpha=e^2/4\pi\sim 1/137$ are almost
always employed, whereas in the context of plasma physics and photon
propagation, unrationalized units corresponding to $\alpha=e^2\sim
1/137$ are used, assuming $\hbar=c=1$ in both cases. We follow the
particle-physics tradition and note that a magnetic field of 1~Gauss
then corresponds to $1.95\times10^{-2}~{\rm eV}^2$. The critical
field strength, defined by $\omega_{\rm c}=m_e$, is then correctly
found as $B_{\rm crit}=m_e^2/e=(0.511~{\rm
MeV})^2/\sqrt{4\pi\alpha}=8.6\times10^{11}~{\rm
eV}^2=4.4\times10^{13}~{\rm Gauss}$.}
\begin{eqnarray}
\omega_{\rm p}&=&e\,\sqrt{\frac{n_e}{m_e}}
=3.7\times10^{-11}~{\rm eV}\,
\left(\frac{n_e}{{\rm cm}^{-3}}\right)^{1/2}\,,
\\
\omega_{\rm c}&=&e\,\frac{B}{m_e}\kern0.9em
=1.16\times10^{-8}~{\rm eV}\,
\left(\frac{B}{{\rm Gauss}}\right)\,,
\end{eqnarray}
where the elementary charge $e$ was taken to be positive. The
analogous contributions from ions are much smaller and will be
neglected.

We decompose ${\bf B}$ into a longitudinal component $B_\parallel$
along the direction of propagation ($z$--direction) and a transverse
one $B_\perp$.  The impact of the magnetic field is determined by
the dimensionless parameters
\begin{equation}\label{eq:bparameters}
b_{\parallel,\perp}=\frac{\omega_{\rm c}}{\omega}\,
\frac{B_{\parallel,\perp}}{B}
=e\,\frac{B_{\parallel,\perp}}{\omega\,m_e}\,.
\end{equation}
Propagating modes exist only if (Appendix A)
\begin{equation}
\omega>{\textstyle\frac{1}{2}}\,\omega_{\rm c}+
\sqrt{{\textstyle\frac{1}{4}}\omega_{\rm c}^2+\omega_{\rm p}^2}\,.
\end{equation}
The assumption that we are dealing with relativistic waves thus
implies that $|b_{\parallel,\perp}|\ll 1$.

The Hamiltonian for the evolution of photon polarization is found to
be (Appendix A)
\begin{equation}\label{eq:ham}
{\sf H}=\frac{\omega_{\rm p}^2}{2\omega}\,
\begin{pmatrix}\vspace{0.2cm}
b_\parallel&-{\textstyle\frac{1}{2}}\,\E^{-\I2\varphi}\, b_\perp^2\\
-{\textstyle\frac{1}{2}} \,\E^{\I2\varphi}\, b_\perp^2&-b_\parallel\end{pmatrix}\,,
\end{equation}
where we have dropped terms proportional to the unit matrix. We have
used the helicity basis, i.e.\ the components of the spinor
$\bm{{\sf A}}$ are the amplitudes of the two circular polarization
components of the wave. The angle $\varphi$ describes the direction
of ${\bf B}_\perp$ relative to a fixed direction transverse to the
photon trajectory.

To linear order in $B$, the matrix ${\sf H}$ is always diagonal and
even if $B$ changes or reverses along the trajectory, nothing new
happens. Photons that begin linearly polarized stay that way, except
that their plane of polarization rotates by an angle proportional to
$\int B_\parallel \D z$. Therefore, we had to include the quadratic
terms, responsible for the Cotton-Mouton effect. It is these terms
that can flip the photon helicity.

The same conclusion is reached by observing that a longitudinal
$B$--field is symmetric relative to rotations around the
$z$--direction. Therefore, angular momentum along $z$ is conserved
and a longitudinal $B$--field can not induce transitions between
photons of opposite helicity. This can be achieved only by the
transverse field. By the same token, transverse fields are required
to achieve mixing with spin-0 particles such neutral pions or axions
or with spin-2 particles such as gravitons.

\subsection{Adiabatic helicity flip}

If photons propagate in the presence of a purely transverse
$B$--field, the Cotton-Mouton effect provides a refractive
difference between the modes that are linearly polarized parallel
and orthogonal to the field. A photon that is initially circularly
polarized will acquire elliptic polarization, eventually reverse its
helicity, and later return to the initial polarization in the spirit
of an oscillation phenomenon.

However, a realistic situation is different. Astrophysical magnetic
fields are not homogeneous and vary in magnitude and direction along
any given photon path. In the relativistic limit where
$|b_{\parallel,\perp}|\ll 1$, the refractive effect of $B_\parallel$
is much larger than that of $B_\perp$ because the Faraday effect is
linear in $B$, the Cotton-Mouton effect quadratic. Therefore, almost
everywhere the ``mass difference'' of photons with opposite helicity
is large compared with the mixing energy provided by the transverse
field. In other words, the unitary transformation between helicity
and propagation eigenstates involves a small mixing angle.

However, $B_\parallel$ may vanish somewhere and may reverse while
$B_\perp$ remains at a typical value. At the $B_\parallel$ inversion
point the helicity eigenstates become degenerate and their effective
masses cross over, except for the small perturbation provided by the
transverse field. It prevents an actual crossing of the eigenvalues
and leads to an adiabatic helicity reversal in the spirit of the MSW
effect.

Within the Landau-Zener approximation~\cite{Landau:1932, Zener:1932ws},
the probability for the state
to \emph{jump} over the avoided level crossing while preserving its
helicity is
\begin{equation}
P_{\rm j}=\E^{-\pi \gamma/2}\,.
\label{eq:pjump}
\end{equation}
The adiabaticity parameter $\gamma$ compares the rate-of-change of
the energy splitting with the oscillation frequency. When the
unperturbed levels would cross (here the $B_\parallel$ reversal),
one finds the usual result
\begin{equation}
\gamma=\frac{4|{\sf H}_{12}|^2}{|{\sf H}'_{22}-{\sf H}'_{11}|}\,
\bigg|_{{\sf H}_{22}={\sf H}_{11}}\,,
\end{equation}
where a prime denotes $\D/\D z$. In our case this is
\begin{equation}
\label{eq:gamma}
\gamma=\frac{\omega_{\rm p}^2}{2\omega}
\,\frac{b_\perp^4}{|2b_\parallel'|}
=\frac{\omega_{\rm p}^2\omega_{\rm c}^3}{4\omega^4}\,\ell_B\,,
\end{equation}
where we have used that at a $B_{\|}$ reversal $B=B_\perp$ and therefore
$b_\perp=\omega_{\rm c}/\omega$. We have also introduced the length
scale of $B$-variation $\ell_B^{-1}=|B_\parallel'/B_\perp|$. If the
magnitude $B$ is fixed so that $B_\parallel=B\cos\theta$ and
$B_\perp=B\sin\theta$ with $\theta$ the angle between photon
direction and ${\bf B}$, we find $\ell_B=|\theta'|^{-1}$. Moreover,
using cycle frequencies $\nu=\omega/2\pi$, we find that the upper
$\nu$ limit given in Eq.~(6) of Ref.~\cite{Broderick:2009ix}
corresponds in our treatment to $\gamma=1$. In other words,
$\gamma=1$ defines a critical frequency
\begin{equation}\label{eq:wcrit}
\omega_{\rm crit}
=\left(\frac{\omega_{\rm p}^2\omega_{\rm c}^3}{4}\,\ell_B\right)^{1/4}
\end{equation}
such that we are in the adiabatic regime for $\omega\ll\omega_{\rm
crit}$ and in the nonadiabatic regime for $\omega\gg\omega_{\rm
crit}$.

The Landau-Zener approximation is strictly applicable only when
${B}_{\|}$ decreases linearly and certain other conditions are met~\cite{Haxton:1986bc}. There is however substantial
literature on its refinements~\cite{Kuo:1989qe}. In particular, the
``double exponential'' ansatz parameterizes the jump probability in
a more widely applicable way~\cite{Petcov:1987zj}. We will continue
to use the Landau-Zener formula for its simplicity, but our
discussion proceeds unaltered if one substitutes a more accurate
expression.

Thus far we assumed that the transverse $B$--field points in a fixed
direction, but of course this direction changes on the same length
scale $\ell_B$ as all other properties. This may have interesting
consequences~\cite{Ralston:1997yp}. However, in the adiabatic
approximation, $B$--field  twisting is slow compared with the
oscillation length and thus would not produce new effects. Rapid
twisting would lead to non-adiabatic transitions.

The $B$--field may have other variations, for example fast
variations caused by turbulence. The evolution of photon
polarization in such circumstances could be performed on the level
of our Schr\"odinger equation. Both for solar and supernova neutrino
oscillations it was found that relatively small density fluctuations
of the medium can severely affect the MSW
effect~\cite{Burgess:1996mz, Fogli:2006xy}. Similar phenomena for
radio waves are expected and would be an interesting subject
of study.

\section{Faraday Rotation}                         \label{sec:faraday}

\subsection{Ordinary Faraday effect}

We now investigate the role of adiabatic helicity flips on the
rotation of linearly polarized radiation traversing a region of
magnetized plasma. If we begin with $\bm{{\sf A}}_0$ at the source,
at the detector we will have
\begin{equation}
\bm{{\sf A}}_{\rm D}={\sf U}_{\rm tot}\bm{{\sf A}}_0\,.
\end{equation}
The unitary matrix taking the initial to the final state is
\begin{equation}
{\sf U}_{\rm tot}={\cal S}\,\exp\left(-\I \int_0^{\rm D}{\sf H}\,\D z\right)\,,
\end{equation}
where the Hamiltonian in the helicity basis is given by
Eq.~(\ref{eq:ham}). The parameters $\omega_{\rm p}$,
$b_{\parallel,\perp}$ and $\varphi$ all depend on $z$. At different
locations the matrices ${\sf H}(z_1)$ and ${\sf H}(z_2)$ do not in
general commute, so the exponential is understood in the
space-ordering convention symbolized by ${\cal S}$. Note that
we work in the helicity basis. However, if photons are produced
and detected in regions that do not involve large transverse fields, the helicity
states are identical to the propagation states at source and detector.
Thus oscillatory terms depending on the mixing angle at source and detector, as
in Ref.~\cite{Parke:1986jy, Petcov:1997am}, will vanish.

In the extreme case when the transverse field $B_\perp$ is either
zero or so small that it plays no role, ${\sf H}$ is always diagonal
and one finds explicitly
\begin{equation}\label{eq:phi}
{\sf U}_{\rm tot}=
\begin{pmatrix} \E^{-\I\,\phi}&0\\0&\E^{+\I\,\phi}\end{pmatrix}
\end{equation}
with the phase
\begin{equation}
\phi=\frac{1}{2\omega}\int_0^{\rm D}\D z\,\omega_{\rm p}^2\,b_\parallel
=\frac{e^3}{2\omega^2m_e^2}\int_0^{\rm D}n_e{\bf B}\cdot\D\bm{\ell}\,.
\end{equation}
The two helicity states pick up equal but opposite phases, implying
that the plane of polarization rotates by the angle $\phi$: the
usual Faraday effect. However, the polarization direction at the
source is not known, so the measurable quantity is the variation of
$\phi$ with frequency. This ``rotation measure'' is
\begin{equation}
{\rm RM}=\frac{\phi}{\lambda^2}=
\frac{e^3}{8\pi^2m_e^2}\int_0^{\rm D}n_e{\bf B}\cdot\D\bm{\ell}\,,
\label{eq:RM}
\end{equation}
where $\omega=2\pi/\lambda$ has been used, with $\lambda$ being the
wavelength. RM itself does not depend on frequency.

\subsection{Adiabatic Faraday effect}

In order to understand the impact of an adiabatic helicity flip we
subdivide ${\sf U}_{\rm tot}$ into several pieces. To be specific we
assume a single $B$-field reversal. The propagation up to somewhat
before this point is described by the ordinary Faraday effect. The
same is true after the reversal onwards. Therefore, the overall
effect is
\begin{equation}
{\sf U}_{\rm tot}={\sf U}_2{\sf U}_{\rm flip}{\sf U}_1
\label{eq:Utot}
\end{equation}
with
\begin{equation}
{\sf U}_{1,2}=
\begin{pmatrix} \E^{-\I\,\phi_{1,2}}&0\\0&\E^{+\I\,\phi_{1,2}}\end{pmatrix}.
\end{equation}
The phases are given by ordinary Faraday integrals on path~1,
leading from the source to the reversal, and path~2, leading from
the reversal to the detector. If the reversal is non-adiabatic,
${\sf U}_{\rm flip}$ is the unit matrix and we recover the previous
result: $\phi=\phi_1+\phi_2$.

On the other hand, if the reversal is perfectly adiabatic it has the
effect of exchanging the left- and right-handed helicity states
\begin{equation}
{\sf U}_{\rm flip}=
\begin{pmatrix} 0&1\\1&0\end{pmatrix}.
\label{eq:adflip}
\end{equation}
In general, an additional relative phase arises---we will be more
precise later. The flip matrix implies
\begin{equation}
{\sf U}_{\rm tot}=
\begin{pmatrix} 0&\E^{\I(\phi_1-\phi_2)}\\ \E^{-\I(\phi_1-\phi_2)}
&0\end{pmatrix}.
\end{equation}
If the initial linear polarization state is $\bm{{\sf A}}_{0}=(1,1)$
the non-adiabatic (na) and adiabatic (ad) final states are
\begin{equation}
\bm{{\sf A}}_{\rm D}^{\rm na}=
\begin{pmatrix} \E^{-\I(\phi_1+\phi_2)}\\ \E^{+\I(\phi_1+\phi_2)}\end{pmatrix},
\quad
\bm{{\sf A}}_{\rm D}^{\rm ad}=
\begin{pmatrix} \E^{+\I(\phi_1-\phi_2)}\\ \E^{-\I(\phi_1-\phi_2)}\end{pmatrix}\,.
\end{equation}
The impact of the second part of the trajectory is the same in both
cases: the helicity components acquire a relative phase $\phi_2$,
whereas the impact of the first part is reversed. The overall
rotation of linear polarization is
\begin{equation}
\Phi^{\rm na}=\phi_1+\phi_2
\quad{\rm and}\quad
\Phi^{\rm ad}=-\phi_1+\phi_2\,.
\end{equation}
In other words, the adiabatic helicity flip has the effect of
reversing the rotation measure accrued on the path before the flip.
This is physically understood: The helicity states in terms of
electric fields are $(E_x\pm \I E_y)/\sqrt2$. Exchanging helicities
amounts to $E_y\to - E_y$ and thus reverses the polarization angle
relative to the $x$--direction.

\begin{figure}
\includegraphics[width=0.8\columnwidth]{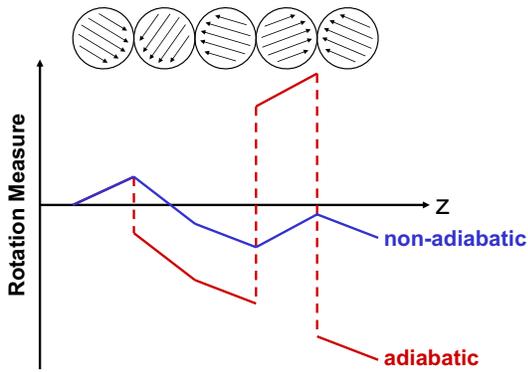}
\caption{Evolution of the rotation measure
in a magnetized plasma with
changing $B$-field orientation. The ordinary (non-adiabatic)
and adiabatic cases are shown.\label{fig:bfields}}
\end{figure}

We juxtapose the non-adiabatic and adiabatic evolution of RM in
Fig.~\ref{fig:bfields} that is analogous to Fig.~2 of
Ref.~\cite{Broderick:2009ix}. An important difference is that for
the adiabatic case these authors actually show the evolution of the
phase difference between the two propagation eigenmodes, a quantity
that indeed always increases. However, the only observable quantity
is the RM. It acquires a minus sign at each adiabatic reversal and
thus jumps by a large amount. The overall adiabatic RM is
\begin{equation}
{\rm RM}^{\rm ad}= \pm\,
\frac{e^3}{8\pi^2m_e^2}
\int_0^{\rm D}n_e|{\bf B}\cdot\D\bm{\ell}|\,.
\label{eq:RMad}
\end{equation}
The absolute sign is identical with the sign of $B_\parallel$ on the
last sub--trajectory.

\subsection{Intermediate adiabaticity}

The final polarization of the beam depends crucially on the degree
of adiabaticity. The entire effect of the reversal is parameterized
in the flip matrix. In general, ${\sf U}_{\rm flip}$ is given by a
unitary matrix
\begin{equation}
{\sf U}_{\rm flip}=\E^{\I \alpha}
\begin{pmatrix} \E^{\I \delta_1}\cos\xi&\E^{\I \delta_2}\sin\xi\;\\
-\E^{-\I \delta_2}\sin\xi&\E^{-\I \delta_1}\cos\xi\;\end{pmatrix},
\end{equation}
where $\cos\xi=\sqrt{P_{\rm j}}$ and ${P}_{\rm j}$ is the
probability to jump between helicities at the crossing. Note the
presence of other phases acquired at the reversal. The non-adiabatic
limit is obtained for $\cos\xi=1$ with other phases set to zero.
Similarly the adiabatic limit is obtained for $\sin\xi=1$, which is
the same as in Eq.~(\ref{eq:adflip}) for overall phase choices
$\alpha=\delta_2=\pi/2$.

If the initial linear polarization state is $\bm{{\sf A}}_{0}=(1,1)$,
the final state (up to an overall phase) is
\begin{equation}
\bm{{\sf A}}_{\rm D}=
\begin{pmatrix}
\E^{\I(\delta_1-\phi_2)}\cos\xi+\E^{\I(\delta_2+2\phi_1-\phi_2)}\sin\xi\\
\E^{-\I(\delta_1-2\phi_1-\phi_2)}\cos\xi-\E^{-\I(\delta_2-\phi_2)}\sin\xi
\end{pmatrix}.
\end{equation}
The intensity of the left and right helicity components of this
state are
\begin{equation}
{I}_{\pm}=1\pm \sin2\xi\cos\left(\delta-2\phi_1\right)\,,
\label{eq:phaseeffect}
\end{equation}
where $\delta=\delta_1-\delta_2$ is a combination of the phases
acquired at the reversal.

At the fully adiabatic or fully non-adiabatic limit, $\sin2\xi=0$
and the two helicities arrive with equal intensity: the beam is
linearly polarized. At intermediate adiabaticity this is no longer
true and the final beam is elliptically polarized to a degree
depending on frequency. Using Eqs.~(\ref{eq:gamma}) and
(\ref{eq:wcrit}), we find
\begin{equation}
\cos\xi=\exp\left[{-\frac{\pi}{4}\left(\frac{\omega}{\omega_{\rm crit}}\right)^4}\right]\;,
\end{equation}
whereas from Eq.~(\ref{eq:phi}) we find that $\phi_1$ varies as
$\int{\D z}\,\omega_{\rm p}^2\,b_{\|}/2\omega$, the integral
extending from the source to the point of reversal.

The degree of elliptical polarization thus has two variations
imprinted upon it as a function of frequency. One arises from the
usual RM accrued between source and field reversal which determines
the orientation with which the linear polarization enters the
cross-over region. The other is a slow variation determined by the
jump probability as a function of frequency, approximately given by
the Landau-Zener formula. If one could measure the polarization
state over a reasonably broad range of frequencies, a measure of
$B$--field tomography would become possible. Unlike the adiabatic or
non-adiabatic limit, at intermediate adiabaticity one has the
opportunity to probe the magnetic fields in specific segments of the
path traveled.

\subsection{Multiple reversals}

We have mostly confined our discussion to a single field reversal,
but the formalism is easily extended to multiple reversals. As
emphasized by Broderick and Blandford, the ordinary Faraday effect
for $N\gg1$ domains, each causing a rotation $\phi_i$, adds up to
\begin{equation}
\Phi^{\rm na}=\sum_{i=1}^{N}\phi_i\propto \sqrt{N}\;.
\end{equation}
On the other hand, the adiabatic Faraday effect leads to a rotation
by
\begin{equation}
\Phi^{\rm ad}=\sum_{i=1}^{N}\phi_i\propto N\;.
\end{equation}
Clearly, for media with multiple reversals this leads to a larger
Faraday rotation at adiabatic frequencies.

For quasi-adiabatic frequencies, multiple reversals lead to loss of
linear polarization by each of the reversals. Schematically, for $N$
domains with $N-1$ field reversals, the intensities of the helicity
components arriving at the detector is
\begin{equation}
\left(
\begin{array}{c}
I_{+}\\
I_{-}
\end{array}
\right)=\prod_{i=1}^{N-1}\left(
\begin{array}{cc}
\cos^2\xi_i&\sin^2\xi_i\\
\sin^2\xi_i&\cos^2\xi_i
\end{array}
\right)
\left(
\begin{array}{c}
1\\
1
\end{array}
\right)\;,
\end{equation}
where we have ignored the fast oscillatory terms. Clearly, there is
increased elliptic polarization at the critical frequencies corresponding to
each of the magnetic field reversal. A more precise
prediction, including interference effects due to the oscillatory
terms, can be calculated using the recipe prescribed in
Ref.~\cite{Dasgupta:2005wn}.

\section{Conclusions}                          \label{sec:conclusions}

We have studied the adiabatic Faraday effect that was recently
discovered by Broderick and Blandford. We have used a simple
formalism that linearizes the photon Klein-Gordon equation and
amounts to a Schr\"odinger equation for a two-level system
consisting of the photon helicities. This approach is commonplace
in the context of neutrino flavor oscillations. Once the photon
dispersion relation has been identified, the adiabaticity condition
follows immediately from well-known textbook results such as the
Landau-Zener approximation. Formulating the adiabatic Faraday effect
in a language familiar from neutrino physics may allow for a broader
appreciation of this intriguing ``MSW effect for photons.''

Despite the similarities, Faraday rotation also has important
differences with neutrino oscillations. For neutrinos, the initial
states are always weak-interaction eigenstates, whereas photons can
in general be produced in any polarization state. During
propagation, neutrinos typically have non-maximal mixing except on
the MSW resonance, causing large flavor transitions. For photons,
the plane of polarization is typically rotated by many full cycles
independently of any transverse field. Therefore, the adiabatic
helicity flip does not enhance the transition between linear
polarization states, it modifies the rotation measure, the way the
rotation of the plane of polarization varies with frequency.
Finally, unlike a neutrino detector, a photon detector in the radio
band has the capability of identifying not merely the intensities of
each polarization state, but also the relative phase between them.
So the observational manifestation of the MSW effect is rather
different for neutrinos and photons.

To summarize, the adiabatic Faraday effect is the process of photon
helicity getting flipped adiabatically in regions of slow $B$-field
reversal, and has the effect of reversing the rotation measure
accrued by the photon up to that point on its trajectory. Therefore,
the subsequent Faraday effect goes effectively in the same direction
as the Faraday effect before the reversal region, giving much larger
overall rotation measure. The global sign of the rotation measure
depends only on the $B$--field direction on the last leg of the
photon path.

The rotation measure both in the adiabatic and non-adiabatic regime
is independent of frequency. Observations over a broad range of
frequencies, ranging from below to above the critical frequency,
would reveal a transition range of fast modulations of elliptic
polarization. In the adiabatic regime, similar variations can arise
in the transverse direction across the sky. The astronomical
potential of these effects for $B$--field tomography has been
explored by Broderick and Blandford. The main purpose of our note
was to clarify the basic principles of the adiabatic Faraday effect
that is a neat application of the formalism usually applied to the
MSW effect in neutrino oscillations.

\section*{Note Added in Proof}

After our manuscript had gone to press, D.~Melrose circulated a
preprint~\cite{Melrose:2010zu}, claiming that the adiabatic
modification of the Faraday effect was not observable. We believe that
underlying this critique is a misunderstanding.

We fully agree with the formalism used by Melrose. Our complex
two-spinor $\bm{{\sf A}}$ is equivalent to a $2\times2$ density matrix
${\sf S}_{ij}=\bm{{\sf A}}_i\bm{{\sf A}}_j^*$ and our Eq.~(\ref{eq:schrodinger}) can be
written as a commutator equation in the form $\I\partial_z{\sf
  S}=[{\sf H},{\sf S}]$. Moreover, any Hermitean $2\times2$ matrix
${\sf R}$ can be written in terms of a three-vector ${\bf R}$ in the
form $\frac{1}{2}{\rm Tr}({\sf R})+\frac{1}{2}{\bf R}\cdot\bm{\sigma}$
with $\bm{\sigma}$ being a vector of Pauli matrices. Our Eq.~(\ref{eq:schrodinger}) is
equivalent to $\partial_z{\bf S}={\bf H}\times{\bf S}$ where ${\bf S}$
is the vector representing ${\sf S}$ and ${\bf H}$ the one
representing ${\sf H}$. Therefore, our Eq.~(\ref{eq:schrodinger}) is equivalent to the
``spin-precession equation'' described after Eq.~(4) of
Ref.~\cite{Melrose:2010zu}. In particular, we agree, that ${\bf S}^2$
is invariant and that ${\bf H}\cdot{\bf S}$ is an adiabatic invariant
if ${\bf H}$ changes slowly as a function of $z$. A precessing spin
following a slowly changing $B$-field is the usual visualization of
adiabatic neutrino oscillations.

\begin{figure}[!t]
\includegraphics[width=0.85\columnwidth]{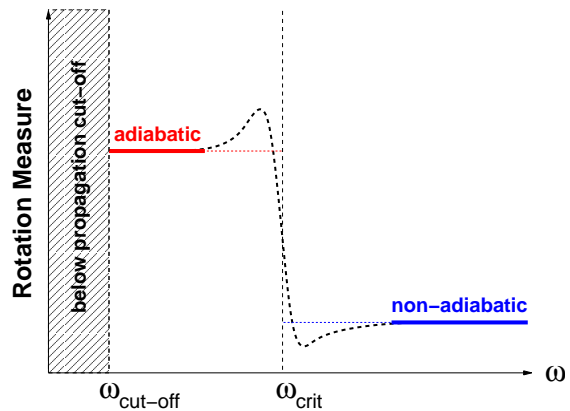}
\caption{The rotation measure as a function of frequency. The ordinary
  (non-adiabatic) and adiabatic regimes are shown.}
\label{fig:fig2}
\end{figure}

The adiabatic limit is defined by the oscillation length being shorter
than the length scale of field reversal, implying that the position
angle (PA) of linearly polarized light must undergo many Faraday
revolutions between source and detector.  Therefore, we agree that PA,
being measurable only modulo $\pi$, carries no information about the
difference between normal and adiabatic Faraday effect. This
information is contained in how quickly PA changes with frequency,
i.e.\ the rotation measure (RM). In our Fig.~\ref{fig:bfields} we have sketched how RM
varies as a function of distance of a hypothetical observer from the
source. However, a real observer is at a fixed location and observes
the effect as a function of frequency as sketched in Fig.~\ref{fig:fig2}. The observed 
RM in the normal and adiabatic regime can be different, and are given 
by Eqs.~(\ref{eq:RM}) and (\ref{eq:RMad}) respectively. Of
course, in the transition region between adiabatic and ordinary
Faraday effect,  the photon polarization is elliptical. Eq.~(\ref{eq:phaseeffect}) 
shows that unless the probability of helicity flip is zero or one, the light is always 
elliptically polarized . Since the RM is not strictly defined in that case, we 
sketch the RM for the major axis of the polarization ellipse.

\section*{Acknowledgements} 

We thank John Beacom and Leo Stodolsky for inspiring discussions and
specific suggestions. This work was partly supported by the Deutsche
Forschungsgemeinschaft under grant TR-27 ``Neutrinos and Beyond''
and the Cluster of Excellence ``Origin and Structure of the
Universe'' (Munich and Garching). B.D.\ thanks the
Max-Planck-Institut f\"ur Physik for support during the initial
stages of this work.

\appendix
\section{Photon dispersion in a cold magnetized plasma}

The propagation of electromagnetic waves in plasma is governed by
Maxwell's equations, including polarizations and currents induced by
the response to the wave. Assuming a homogeneous cold magnetized
plasma, harmonic time variation $\E^{-\I\omega t}$ for all
quantities, and spatial variation of the wave only along the
$z$--direction, one finds a stationary Klein-Gordon equation for the
electric field vector ${\bf E}$ of the form~\cite{Ginzburg:1970}
\begin{equation}\label{eq:KGE}
-\begin{pmatrix}\partial_z^2\\&\partial_z^2&\\&&0\end{pmatrix}
{\bf E}=\left(\omega^2-\omega_{\rm p}^2\,\hat\Pi\right){\bf E}\,,
\end{equation}
where $\omega_{\rm p}$ is the plasma frequency. $\hat\Pi$ is the
reduced polarization tensor (polarization tensor in units of
 $\omega_{\rm p}^2$) which depends on $\omega$ and the medium
properties.

Assuming the ${\bf B}$ field has components $B_z=B_\parallel$,
$B_x=B_\perp$ and $B_y=0$ and using the $b$ parameters defined in
Eq.~(\ref{eq:bparameters}) one finds
\begin{equation}
\hat\Pi=
\dfrac{1}{1-b_\parallel^2-b_\perp^2}\,
\left(
\begin{array}{ccc}
\vspace{0.2cm}
1-b_{\perp}^2&
\I b_{\|}&
-b_{\|}b_{\perp}\\
\vspace{0.2cm}
-\I b_{\|}&1&
\I b_{\perp}\\
-b_{\|}b_{\perp}&
-\I b_{\perp}&
1-b_{\|}^2\\
\end{array}
\right)\,.
\end{equation}
The lowest frequencies arise in the homogeneous case when all
spatial derivatives vanish and the l.h.s.\ of Eq.~(\ref{eq:KGE})
vanishes identically, leading to
\begin{equation}
\omega_0=\omega_{\rm p}
 \quad\hbox{or}\quad
\omega_0=
{\textstyle\frac{1}{2}}\,\omega_{\rm c}+
\sqrt{{\textstyle\frac{1}{4}}\omega_{\rm c}^2+\omega_{\rm p}^2}\,.
\end{equation}
The first solution corresponds to the ordinary plasma oscillation.
The other connects to the dispersion relation for propagating modes
for non-vanishing $k$ and thus is the minimum frequency required for
a propagating wave.

For propagating modes the $z$--equation of Eq.~(\ref{eq:KGE})
represents a constraint, allowing one to eliminate $E_z$. We
represent the propagating modes by the usual Jones vector $\bm{{\sf
A}}=(E_x,E_y)$, leading to a stationary Klein-Gordon equation in the
form
\begin{equation}\label{eq:KG2}
-{\partial}_z^2\, \bm{{\sf A}}=
\left({\omega^2}-\omega_{\rm p}^2\hat{\sf\Pi}\right)
\bm{{\sf A}}\,.
\end{equation}
The $2\times2$ reduced polarization matrix is found to be
\begin{equation}
\hat{\sf\Pi}_{ij}=\hat{\Pi}_{ij}+ \frac{\hat{\Pi}_{iz}\hat{\Pi}_{zj}}
{\omega^2/\omega_{\rm p}^2-\hat{\Pi}_{zz}}\,,\quad
i,j=x~{\rm or}~y\,.
\end{equation}
With the dimensionless parameter $b_{\rm p}=\omega_{\rm p}/\omega$
we find for $\hat{\sf\Pi}$ the expression
\begin{equation}\label{eq:poltensor2}
\frac{1}{\big(1-b_{\rm p}^2\big)\big(1-b_\parallel^2\big)-b_\perp^2}
\begin{pmatrix}\vspace{0.2cm}
1-b_{\rm p}^2-b_\perp^2&\I b_\parallel\big(1-b_{\rm p}^2\big)\\
-\I b_\parallel\big(1-b_{\rm p}^2\big)&1-b_{\rm p}^2\end{pmatrix}.
\end{equation}
Notice that the absence of absorption renders this matrix Hermitean.

If all of $b_{{\rm p},\perp,\parallel}$ are small compared to unity
and we expand up to quadratic order we find
\begin{equation}\label{eq:poltensor3}
\hat{\sf\Pi}=1+b_{\parallel}^2+
\begin{pmatrix}\vspace{0.2cm}
0&\I b_\parallel\\
-\I b_\parallel&b_{\perp}^2\end{pmatrix} +{\cal O}(b^3)\,.
\end{equation}
In general the ${\bf B}$--field has components $B_\perp\cos\varphi$,
$B_\perp\sin\varphi$ and $B_\parallel$ relative to a coordinate
system with a fixed $x$--direction, implying that the polarization
matrix must be rotated correspondingly. Moreover, for some purposes
it is simpler to work in the helicity basis where the electric field
has the two components $(E_x\pm\I E_y)/\sqrt{2}$. Altogether these
transformations lead to
\begin{equation}\label{eq:poltensor4}
\hat{\sf\Pi}=1+b_{\parallel}^2+{\textstyle\frac{1}{2}}b_{\perp}^2+
\begin{pmatrix}\vspace{0.2cm}
b_\parallel&-{\textstyle\frac{1}{2}}\,\E^{-\I2\varphi}\, b_\perp^2\\
-{\textstyle\frac{1}{2}} \,\E^{\I2\varphi}\, b_\perp^2&-b_\parallel\end{pmatrix}\,.
\end{equation}
This is the result used in the main text.


\end{document}